\documentclass[aps,prl,twocolumn,groupedaddress]{revtex4}

\usepackage{graphicx}
\usepackage{bm}

\begin{document}
\title{Experimental Realization of an Exact Solution to the Vlasov Equations for an Expanding Plasma}
\author{S. Laha$^1$, P. Gupta$^1$, C. E. Simien$^1$, H. Gao$^1$,
J. Castro$^1$,  T. Pohl$^2$, and T. C. Killian$^1$}
\affiliation{$^1$Rice University, Department of Physics and
Astronomy,
Houston, TX, USA 77005\\
$^2$ITAMP, Harvard-Smithsonian Center for Astrophysics, 60 Garden
Street, Cambridge, MA, USA 02138}
\date{\today}

\begin{abstract}
We study the expansion of ultracold neutral plasmas in the regime in
which inelastic collisions are negligible.
%into a surrounding vacuum
%from the early acceleration phase until the onset of terminal
%velocity.
The plasma expands due to the thermal pressure of the electrons, and
for an initial spherically symmetric Gaussian density profile, the
expansion is self-similar. Measurements of the plasma size and ion
kinetic energy using fluorescence imaging and spectroscopy show that
the expansion follows an analytic solution of the Vlasov equations
for an adiabatically expanding plasma.

\end{abstract}

%\pacs{32.80.Pj}
%\keywords{}

\maketitle

Exactly solvable problems are rare in physics and serve as ideal
models that provide a starting point for understanding more complex
systems. Here, we report the experimental realization  of a
laser-produced plasma whose dynamics can be described by an exact
analytic solution to the Vlasov equations \cite{bku98,dse98}, which
are central equations in the kinetic theory of plasmas. Expansion
into a surrounding vacuum is fundamentally important and typically
dominates the dynamics of plasmas created with pulsed lasers
\cite{pmo94}, such as in experiments pursuing inertial confinement
fusion \cite{lin95}, x-ray lasers \cite{dai02}, or the production of
energetic ($>$MeV) ions through irradiation of solids
\cite{ckz00,skh00}, thin foils
\cite{mgf00,bwp01,msp02,hkp02,kss04,fsk05,rfb05,mkn05}, %asa03,hbc00
clusters \cite{shh07}. %dts97}.%, and gas jets \cite{sbn99,kcn99}.
%Intense ion beams have great promise for medical applications \cite{szb01,kmi98}, for triggering nuclear reactions \cite{btt99},
%or as compact neutron sources \cite{psp98,nfb98,dzy99,dgm99}.

We study  plasma expansion with ultracold neutral plasmas (UNPs) \cite{kkb99,kil07},
which are created by
photoioinizing laser-cooled strontium atoms \cite{nsl03} just above the ionization threshold. These systems
occupy a new regime of neutral plasma physics, and their well-controlled initial conditions and relatively slow
dynamics have enabled precise studies of strongly coupled plasma properties \cite{scg04,csl04}. Here we use
fluorescence imaging and spectroscopy for the first time on these systems. These powerful diagnostics allows us
to apply the advantages of UNPs to a new class of important problems.

The investigation of plasma expansion dates back many decades
\cite{gpp66, ssc87}.
%Older studies often assumed
%isothermal electrons \cite{ssc87}.
%% because of the relevance to
%%dense plasmas in strong thermal contact with a solid target.
%However, adiabatic cooling of the electrons must be taken into
%account for long expansion times in all systems \cite{lro86,mmf93}
%and even for early times for plasmas created from thin foils,
%clusters, or gases \cite{bku98,dse98,mor05pre}. As a result, much
%recent theoretical work has focused on expansion
%\cite{mor05physplasmas,mor05pre,kby03,ccb05,mkn05} when electrons
%cool adiabatically. Theory \cite{mkn05} matches experimental
%measurement of the ion-energy spectrum resulting from irradiation
%of a spherical thin-film target and a gas jet, and general scaling
%trends of cluster explosion energy  have been found \cite{sht00}.
%However, to the best of our knowledge, a complete and consistent
%experimental and theoretical characterization of ion and electron
%expansion dynamics has not been shown.
Recently,  exact solutions for spatially finite plasmas expanding
into vacuum were  identified for a 1-dimensional plasmas
\cite{bku98} and later extended to 3-dimensions  \cite{dse98,kby03}.
This work was motivated by plasmas produced with short-pulse lasers.
%
%The Vlasov equations for the electron ($\alpha=e$) and ion
%($\alpha=i$) distribution functions are
%\begin{equation} \label{vlasov}
%\frac{\partial f_\alpha}{\partial t}+\textbf{v}\cdot
%\nabla_\textbf{r}
%f_\alpha-\frac{e}{m_\alpha}\nabla_\textbf{r}\varphi \cdot
%\nabla_\textbf{v}f_\alpha=0,
%\end{equation}
%where the electrostatic potential is given by Poisson's equation
%\begin{equation} \label{poissons}
%\nabla^2_\textbf{r}\varphi=e(n_e- n_i)/\varepsilon_0,
%\end{equation}
%and the electron and ion densities are
%\begin{equation} \label{densities}
%n_\alpha(\textbf{r})=\int d\textbf{v}
%f_\alpha(\textbf{r},\textbf{v}).
%\end{equation}

The Vlasov equations, along with Poisson's equation, describe the
evolution of the electron ($\alpha=e$) and ion ($\alpha=i$)
distribution functions, $f_\alpha(\textbf{r},\textbf{v})$.
 The Vlasov equations
neglect radiative processes and collisional phenomena
 such as  electron-ion thermalization and three-body
 recombination \cite{mke69}, but this
 formalism describes many types plasmas and
is part of the foundation of kinetic theory.
%Even if collisions are
%important for maintaining equilibrium for each species, the dynamics
% is typically called collisionless
%because a Fokker-Plank term vanishes for a Maxwell-Boltzmann
%distribution function \cite{bku98}.

Among broad classes of general analytic solutions to the
Vlasov
equations \cite{kby03}, UNPs realize a particular solution,
that
even describes collisional systems and is valid for a
quasi-neutral
plasma with spherical Gaussian distribution functions
\begin{equation} \label{gaussianequation}
f_\alpha\propto\mathrm{exp}\left[-\frac{r^2}{2\sigma^2}
-\frac{m_\alpha(\mathbf{v}-\mathbf{u})^2}{2k_\mathrm{B}
T_\alpha}\right].
\end{equation}
Quasi-neutrality is defined by $n_e\approx n_i$, where the electron
and ion densities are $ n_\alpha(\textbf{r})=\int d\textbf{v}
f_\alpha(\textbf{r},\textbf{v})$. $T_\alpha$ are the electron and
ion temperatures, and
 the
local average velocity  varies in space according to ${\bf u}({\bf
r},t)=\gamma(t){\bf r}$.
% This
%produces an electric field that increases linearly with distance
%from the origin, which causes a self-similar expansion that
%preserves the plasma shape.
The temperatures must scale as $\sigma^2T_\alpha=constant$
\cite{bku98}, which is expected for adiabatic cooling in a
spherically symmetric UNP \cite{rha03,ppr04PRA}.
%(Note: In the absence of spherical symmetry, an
%analytic solution is still possible, but it requires spatially
%anisotropic temperature distributions.)

Under these conditions, the plasma dynamics is given by
%\begin{equation} \label{eq2}
%\dot{\sigma}=\gamma\sigma\;,\quad \dot{T}_\alpha=-2\gamma
%T_\alpha\;,\quad\dot{\gamma}=\frac{k_{\rm B}(T_{\rm e}+T_{\rm
%i})}{m_{\rm i}\sigma^2}-\gamma^2.
%\end{equation}
%Equation~(\ref{eq2}) yields
$T_\alpha(t)=T_\alpha(0)/(1+t^2/\tau_{\rm exp}^2)$, where the
characteristic expansion time $\tau_{\rm exp}$ is given by
$\tau_{\rm exp}=\sqrt{m_{\rm i}\sigma(0)^2/k_{\rm B} [T_{\rm
e}(0)+T_{\rm i}(0)]}$. Also,
\begin{equation}\label{eqexpansion}
    \sigma(t)^2=\sigma(0)^2(1+t^2/\tau_{\rm exp}^2),
\end{equation} and
\begin{eqnarray} \label{eq3}
\mathrm{\textsl{v}}_{i,rms} = \sqrt{\frac{k_{\rm B}}{m_{\rm
i}}\left\{\frac{t^2}{\tau_{\rm exp}^2}\left[T_{\rm e}(t)+T_{\rm
i}(t)\right]+T_i(t)\right\}}\;
\end{eqnarray}
describe the evolution of the characteristic plasma size and  ion
velocity, which are important experimental observables. We define
the rms 1-dimensional ion velocity $\sqrt{\langle(\textbf{v}\cdot
\textbf{\^{y}})^2\rangle } \equiv \mathrm{\textsl{v}}_{i,rms}$,
where $\textbf{\^{y}}$ is the laser propagation direction,
$\mathbf{v}$ is the total ion velocity including random thermal
motion and expansion, and the angled brackets refer to an average
over the
 distribution function.

%Qualitatively, the expansion dynamics is similar in all quasineutral
%laser-produced plasmas.
%and it was described in early work \cite{ply60,hre62}.
%A small fraction of the electrons, which have been heated  and
%perhaps accelerated by the laser, leave the plasma and enter the
%surrounding vacuum, creating a strong ambipolar electric field that
%accelerates the ions outward.
%Theory for thin foil experiments shows that the peak accelerating field occurs in the electrostatic sheath at
%the rear of the target and is $E \approx k_\mathrm{B} T_e/e\lambda_D$ \cite{hbc00,gmo06}, where $T_e$ is the electron temperature,
%$e$ is the fundamental charge, and $\lambda_D=(\varepsilon_0 k_\mathrm{B} T_e/ n_{e} e^2)^{1/2}$ is the Debye length for electron density $n_e$.
%In this quasineutral system, the mean electric field can be
%expressed as $-k_\mathrm{B} T_e \nabla n_i/en_i$ \cite{ppr04}, which in a
%Gaussian system leads to a force on the ions of
%\begin{equation}\label{force}
%    \vec{F}(t)={k_\mathrm{B} T_e (t) \vec{r} \over \sigma(t)^2}.
%\end{equation}
%Linearity with $\vec{r}$ gives rise to a self-similar expansion.
An intuitive explanation for the self-similar nature of the
expansion is that  thermal pressure produces an average radial
acceleration \cite{dse98,ppr04PRA}
\begin{equation} \label{eq1}
\dot{\bf u}=-\frac{k_{\rm B}\left[T_{\rm e}(t)+T_{\rm
i}(t)\right]}{m_{\rm i}}\frac{\nabla n(r,t)}{n(r,t)}=\frac{k_{\rm
B}\left[T_{\rm e}(t)+T_{\rm i}(t)\right]}{m_{\rm i}\sigma(t)^2}{\bf
r}\;.
\end{equation}
The simplification implied by the last equality is only valid for a
spherical Gaussian plasma, and the linearity in $\textbf{r}$
preserves the shape of the distribution functions.

Plasmas produced with solid targets, foils, and clusters are often
quasi-neutral and well-described by the Vlasov equations, and
electrons cool adiabatically during much of their evolution, but
experimental conditions studied are typically very complicated and
evolve extremely rapidly, which frustrates detailed comparison
between experiment and theory. Final ion kinetic energy
distributions have been shown to agree with simple models
\cite{mkn05,shh07},
% based on a
%hydrodynamic expansion under appropriate conditions . %or
%Coulomb explosion \cite{shh07}.
but in general, these systems lack the Gaussian distribution
functions necessary to realize the analytically describable
self-similar expansion.% \cite{mor05pre}.

For appropriate initial conditions, UNPs \cite{kkb99} fulfill the
requirements for the analytic solution. UNPs have ion temperatures
of about 1\,K determined by disorder-induced heating
 \cite{mur01,scg04}. Electron
temperatures ranging from 1 to 1000\,K are set by the detuning,
$E_e$, of the ionization laser above threshold. The peak densities
are on the order of $10^{15}$\,m$^{-3}$, and the profile follows
that of the laser-cooled atom cloud, which we adjust to have a
spherically symmetric Gaussian. The photoionization pulse length
($\sim 10$\,ns) is much less than the expansion time scale ($\sim
10$\,$\mu$s).

%This analysis assumes spherical symmetry, but the raw plasma images
%show that $\sigma_x=\sigma_y$ within 5\%, and
%$\sigma_x=\sigma_y=\sigma_z$ is initially established by imaging the
%laser-cooled atom cloud along orthogonal directions.

 The electron distribution equilibrates locally
within 100\,ns and globally within 1\,$\mu$s after photoionization
\cite{rha03,ppr04PRA}. This ensures a Gaussian electron distribution
function at the start of the expansion. %(Note: Truncation effects
%due to the finite depth of the Coulomb potential trapping the
%electrons are negligible for this study \cite{ppr04PRA}.)
Despite these very rapid electron-electron collisions the
corresponding collision integral vanishes for the spherically
symmetric, Gaussian velocity distribution eq.(1). Hence the highly
collisional UNP considered here provides an ideal model system for
truly collisionless plasmas behavior.

 Ions reach local thermal
equilibrium within a few 100\,ns \cite{scg04}. They do not
equilibrate globally on the time scale of the expansion
\cite{csl04}, but the ions are so cold compared to the electrons,
that the lack of a global ion temperature does not cause any
significant deviation from the exact solution. The low ion
temperature also implies that the ions form a strongly coupled fluid
\cite{ich82,scg04}, which, however, negligibly affects the plasma
expansion \cite{ppr04PRA}.

%, and absorption imaging
%and spectroscopy of the plasma \cite{scg04} provide  powerful
%diagnostics of the ion  spatial and velocity distributions.

The analytic expansion solution has been discussed previously for
UNPs \cite{rha02,rha03,ppr04PRA},
 %(Note: Effects due to strong-coupling of the ions  were found to be small
%\cite{ppr04PRA} and will be neglected in this study.),
and it has been checked against average terminal ion expansion
velocities \cite{kkb00} and measured electron temperatures
\cite{rfl04} that qualitatively affirm the importance of adiabatic
cooling for appropriate initial conditions. The lack of spatial and
temporal resolution, available here, however prevented conclusive
tests of the analytic predictions. Cummings \textit{et al.}
\cite{cdd05,cdd05physplasmas} adapted the formalism of
\cite{rha02,rha03,ppr04PRA} and used light-scattering from a small
region of the plasma to study the expansion of ultracold plasmas
with an elongated aspect ratio, but they found significant
deviations from the predictions of the model that perhaps arose
because the condition of spherical symmetry
%-required for self-similar expansion and proper scaling of the electron temperature in the collisional regime-
was not fulfilled.

 To demonstrate that ultracold neutral plasmas can realize the
 analytic expansion solution,  we will first describe our diagnostic and show that the plasma remains
 Gaussian during its expansion. Then we
will show the size variation and ion velocity are given by Eqs.\
\ref{eqexpansion}  and \ref{eq3} respectively. %(ion and electron
%temperature?)
%which, through its dependence on electron temperature and size,
%confirms Eqs.\ \ref{expansion}eqexpansion and \ref{eqn:Tewithtime}.

\begin{figure}[ht]
%\centering
%\includegraphics[clip=true,angle=-90,width=4in, trim=50 50 0 125]{fluorescencesetup.eps}
\includegraphics[clip=true,angle=-90,width=3.25in]{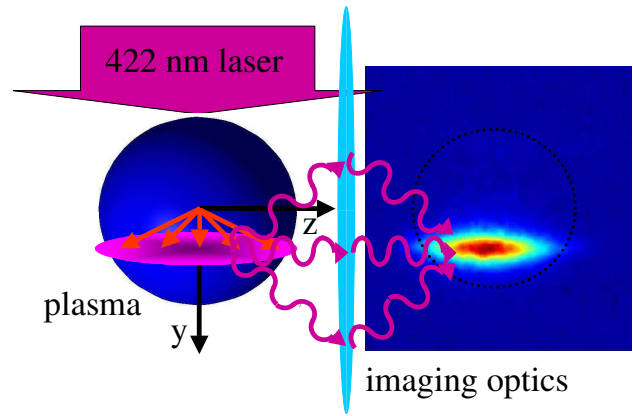}%\includegraphics[clip=true,angle=0,width=3in]{//strontium/Strontium86/PlasmaExpansionPaper/Self-similar-gaussian-paper/gaussianannularfigures/selectedfigures/spectrum3d+rawimage1.eps}
\caption{\label{fluorescenceprobe} Recording fluorescence of UNPs.
The correlation between position and expansion velocity (red arrows)
produces a striped image when the Doppler-shift due to expansion
exceeds the Doppler shift associated with thermal ion velocity.}
\end{figure}

 Figure \ref{fluorescenceprobe} shows a schematic of the
 fluorescence imaging experiment. A laser beam that is near resonance with
 the $^2S_{1/2}-{^2P_{1/2}}$ transition in Sr$^+$ at $\lambda=422$~nm
 propagates along $\hat{y}$ and
 illuminates the plasma. Fluorescence in a perpendicular
 direction ($\hat{z}$) is imaged with a 1:1 relay telescope onto an
 image-intensified CCD camera. The $422$~nm light is typically applied in a 1\,$\mu$s pulse
to provide temporal resolution, and the intensity is only a few
mW/cm$^2$, which is low enough to avoid optical pumping to the
metastable $^2D_{3/2}$ state.

A general expression for the fluorescence is
\begin{eqnarray} \label{eqn:spectrum}
&&F(\nu,x,y) \propto \int ds
\frac{1}{1+\left[\frac{2(\nu-s)}{\gamma_{eff}/2\pi}\right]^2}\times  \nonumber \\
&&\hspace{-.8cm} \int\hspace{-.2cm}\frac{dz\,
n(\mathbf{r})}{\sqrt{2\pi}\sigma_D[T_{i,therm}(\mathbf{r})]}{\rm{exp}}\left\{-\frac{\left[s-(\nu_0+\nu_{exp}^y(\mathbf{r}))\right]^2}{2\sigma_D^2[T_{i,therm}(\mathbf{r})]}\right\},
\end{eqnarray}
where $\nu$ is the laser frequency and $\gamma_{eff}=\gamma_{0} +
\gamma_{laser}$ is the sum of the
 natural linewidth of the transition ($2\pi\times 20$\,MHz) and the
imaging laser linewidth ($2\pi\times 8$\,MHz).
$T_{i,therm}(\mathbf{r})$ is the local temperature of the ions
describing random thermal motion, which gives rise to the Doppler
width $\sigma_D$. Due to the directed expansion velocity, the
average resonance frequency of the transition for atoms at
$\mathbf{r}$  is Doppler-shifted from the value for an ion at rest,
$\nu_0$, by $\nu_{exp}^y(\mathbf{r})=\mathbf{u}(\mathbf{r})\cdot
\hat{y}/\lambda$. The spatial variation in $T_{i,therm}$, which
varies as $n_i^{1/3}$ \cite{csl04}, is small compared to the
directed expansion energy, so $T_{i,therm}$ can be taken as
constant.
 %(only a few hundred mK
%for the central $r\,$\raisebox{-.6ex}{$\stackrel{<}{\sim}$}$\,2
%\sigma$, that contributes most strongly to the signal.)
%Thus, in
%Eq.\ \ref{eqn:spectrum}, $T_{i,therm}$ can be assumed constant in
%space and the density integral can be solved.

Images can be analyzed in several different ways, which each provide
access to different plasma properties. Summing  a series of images
taken at equally spaced frequencies covering the entire ion
resonance is equivalent to integrating $F(\nu,x,y)$ over frequency.
This yields a signal proportional to the areal plasma density, $\int n(x,y,z)dz$, which
for a Gaussian density distribution should take the form
$n_{areal}(x,y)= \sqrt{2\pi}\sigma
n_0\mathrm{exp}\left[-(x^2+y^2)/{2\sigma^2}\right]$.

\begin{figure}
\includegraphics[clip=true,angle=-90,width=3.25in]{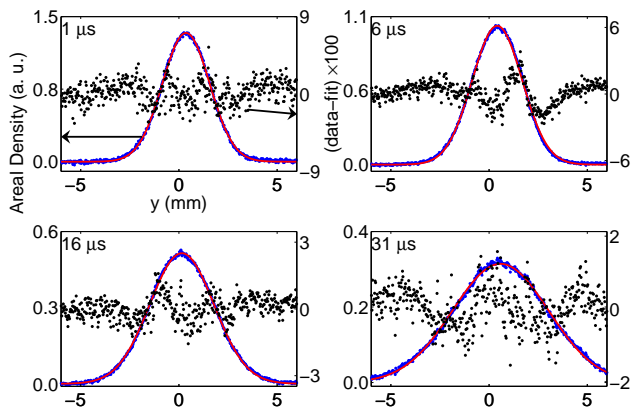}
\caption{ Evolution of the areal density for a plasma with $1.6\pm
0.1\times 10^8$ ions, $2E_e/3k_\mathrm{B}=48\pm3$\,K, and
$\sigma(0)=1.3\pm 0.1$\,mm. The arial density (arbitrary units) is
found by summing together 50 images taken at equally spaced
frequencies that fully cover the ion resonance. The Gaussian fits
(solid line) to
 linear cuts show that the expansion is self-similar. The time
indicated is the evolution time since plasma creation. The
right-hand axes show that the differences between data and fit are
small.} \label{gaussian}
\end{figure}

Figure \ref{gaussian} shows that Gaussian fits of the areal density
are excellent during more than  a factor of two change in $\sigma$
from the earliest times until the signal expands beyond the range of
the imaging system. This provides direct confirmation of the
self-similar nature of the expansion.

There is no sign of any deviation from the gaussian shape at large
radius. This might seem surprising because self-similarity follows
from Eq.\ \ref{eq1}, and this equation must break down at large $r$
where it implies unphysical accelerations arising from
unphysically large electric fields. %As pointed out in
%\cite{ssc87,rha02,mor05physplasmas},
One would expect a breakdown where the plasma is not quasineutral,
which should occur when  the local Debye screening length exceeds
the length scale for ion density variation, or $\sigma \ll
\lambda_D (r)=[\varepsilon_0 k_\mathrm{B} T_e/ n_{e}(r) e^2]^{1/2}$
for a Gaussian plasma. At the radius where $\sigma \approx \lambda_D
(r)$, the accelerating electric field reaches a maximum of
$\sim$$k_\mathrm{B}T_e/e\lambda_D$, identical to peak accelerating
fields in plasmas generated from foil targets \cite{gmo06}. A peak
in the electric field can lead to wave-breaking, shock waves, and
the formation of an ion front as seen in theories for
 UNPs \cite{rha02,rha03,ppr04PRA} and traditional short-pulse
 laser-produced plasmas \cite{mor05physplasmas,mkn05,kds03}.
The lack of such
 features  in data such as Fig.\ \ref{gaussian}
 confirms the
  extent of quasineutrality one would expect for such a low
  electron temperature \cite{rha02,rha03}
and may also
 suggest that finite
ion temperature and strong coupling effects damp wave-breaking at
the plasma edge \cite{ppr04PRA}.

 %Eq.\ \ref{eq1} assumes quasineutrality, which only
%exists if the length scale for ion density variation exceeds the
%local Debye screening length, or $\sigma \gg \lambda_D
%(r)=[\varepsilon_0 k_\mathrm{B} T_e/ n_{e}(r) e^2]^{1/2}$ for a
%Gaussian plasma. The low electron temperature in UNPs ensures
%quasineutrality  up until
% a radius of
% a few $\sigma$ \cite{rha02}. At this radius the accelerating electric field
%reaches a maximum of $\sim k_\mathrm{B}T_e/e\lambda_D$, identical to
%peak accelerating fields in plasmas generated from foil targets
%\cite{gmo06}. A peak in the electric field can lead to
%wave-breaking, shock waves, and the formation of an ion front as
%seen in theories for
% UNPs \cite{rha02,rha03,ppr04PRA} and traditional short-pulse
% laser-produced plasmas \cite{mor05physplasmas,mkn05,kds03}. The lack of such
% features  in images such as Fig.\ \ref{gaussian}
% confirms the extent of quasineutrality of the plasma
%and may
% suggest that finite
%ion temperature and strong coupling effects damp wave-breaking at
%the plasma edge \cite{ppr04PRA}.

\begin{figure}[t]
\centering
\includegraphics[clip=true,angle=270,width=3.25in, trim=0 0 0 0]{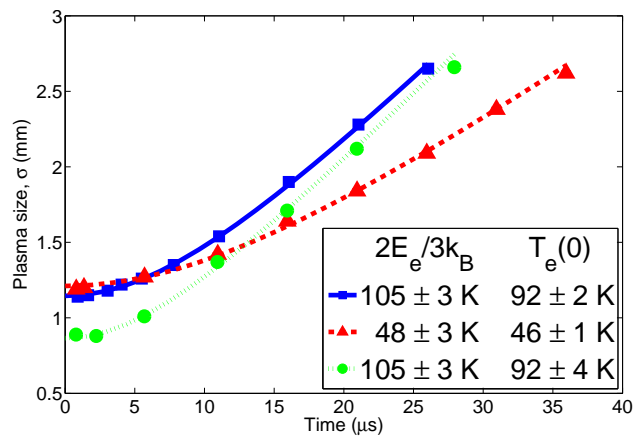}
\caption{\label{figexpansion} Evolution of the plasma size,
extracted from fits such as in Fig.\ \ref{gaussian}.  $T_e(0)$ and
$\sigma(0)$ are fit, while $T_i(0)$ is taken from the theoretical
expression for disorder-induced heating \cite{mur01,scg04}.
Uncertainties in $2E_e/3k_\mathrm{B}$ reflect 1-standard-deviation
calibration uncertainty in the wavelength of the photoionizing
laser. Quoted uncertainties in $T_{e}(0)$ are statistical, but there
is an additional systematic uncertainty of a few percent arising
from calibration of the imaging-system magnification and overlap of
the plasma and fluorescence excitation laser. Statistical
uncertainty in the measurement of $\sigma$ is less than the size of
the plotting symbols. Initial peak densities for these samples are
$\sim 10^{16}\,\rm{m}^{-3}$.}
\end{figure}

The evolution of $\sigma$ can be extracted from fits such as Fig.\
\ref{gaussian}, as shown in Fig. \ref{figexpansion}  for several
different plasma conditions. Again, the data evolve as predicted by
theory. The expansion is sensitive to the electron temperature and
initial size, and higher electron temperature and smaller  size lead
to a faster expansion because this increases the thermal pressure
(Eq.\ \ref{eq1}). Fits of the data using Eq.\ \ref{eqexpansion}
yield values of $T_{e}(0)$ that agree reasonably well with the
expected values of $2E_e/3k_\mathrm{B}$. This confirms that on the
time scale of the expansion there are no significant collisional or
radiative processes changing the electron temperature such as
three-body recombination or electron-ion thermalization, as assumed
in the Vlasov equations. %At higher initial plasma density and lower
%$E_e$, this is not the case \cite{kkb00,kon02,mck02,rha03}. The
%transition to a  plasma whose electron temperature evolution is
%dominated by inelastic collisional processes will be the subject of
%a future study.

\begin{figure}[t]
\centering
\includegraphics[clip=false,angle=270,width=3.25in]{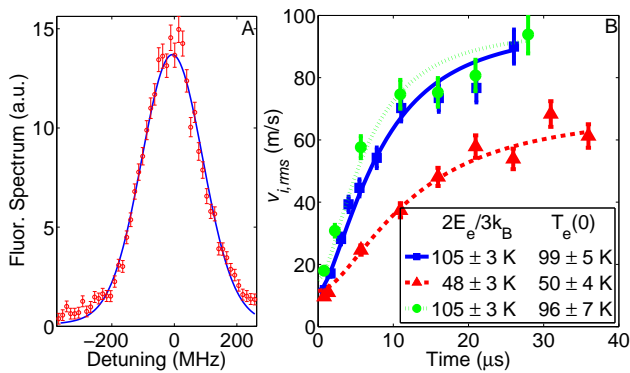}
\caption{\label{figrmsvelocity} (A) Sample spectrum and (B)
evolution of the rms ion velocity ($\textsl{\textrm{v}}_{i,rms}$).
(A) Spectrum  for $2E_e/3k_\mathrm{B}=105 \pm 3$\,K, peak ion
density of $3\times 10^{15}\,\rm{m}^{-3}$, and plasma evolution time
of  3\,$\mu$s. Fits to a Voigt profile provide
$\textsl{\textrm{v}}_{i,rms}$. (B) The velocity evolution
  is fit well by  Eq.\ \ref{eq3}
 with $T_{e}(0)$  as a fit
parameter, $\sigma(0)$ determined from the cloud size measurements
(Fig.\ \ref{figexpansion}), and $T_{i}(0)$ taken from the
theoretical expression for disorder-induced heating
\cite{mur01,scg04}. }
\end{figure}

In order to completely characterize the plasma expansion we also
measure the light-scattering resonance spectrum, formed from the
integrated fluorescence in each of a series of images taken at
different frequencies, as shown in  Fig.\ \ref{figrmsvelocity}A.
Eq.\ \ref{eqn:spectrum}, combined with the expansion velocity
$\textbf{u}(\textbf{r})$ predicted by the Vlasov equations, implies
that the resulting signal ($\int dx\,dy\,F(\nu,x,y)$) should take
the form of a Voigt profile. The rms width of the Gaussian component
of this profile arising from Doppler broadening reflects both
thermal ion motion and directed expansion and is given by
$\tilde{\sigma}_D=\textsl{\textrm{v}}_{i,rms}/\lambda$.

 Equation \ref{eq3} provides an excellent fit to
the data, and the extracted values of $T_{e}(0)$ are consistent with
$2E_e/3k_\mathrm{B}$ within experimental uncertainty. The small
initial offset of $\textsl{\textrm{v}}_{i,rms}$ in each plot is due
to disorder-induced
 heating of the ions within the first microsecond \cite{scg04}, which locally produces
 a thermal ion velocity distribution. But $\textsl{\textrm{v}}_{i,rms}$
 quickly increases well above this value as the electron pressure
drives the plasma expansion and electron thermal energy is converted
into directed radial ion velocity.
 As $\sigma$ increases and the electrons cool
 adiabatically, the acceleration decreases (Eq.\ \ref{eq1}) and the
 ions eventually reach terminal velocity when essentially all electron kinetic
energy is transferred to the ions. %The terminal velocity was
%measured in \cite{kkb00} and first described theoretically in
%\cite{rha02}. Spatially resolved  fluorescence detection of the ions
%\cite{cdd05,cdd05physplasmas}  in a cylindrical plasma measured
%expansion energies  similar to what was observed in \cite{kkb00},
%but they found significant deviations from theoretical predictions
%that perhaps arose because of the lack of spherical symmetry.

Our measured density profiles (Fig.\ \ref{figexpansion}) confirm the
validity of Eq.\ \ref{eq3}, which shows that the ion acceleration is
sensitive at all times to the instantaneous electron temperature and
width of the plasma. Agreement between experiment and theory for
both the size evolution and the ion velocity dynamics hence
demonstrates that $T_{\rm e}(t)$ also follows the dynamics predicted
by the analytical solution to the Vlasov equations.

%Should we show measurements of the ion temperature in the center of
%the cloud that show that the ion temperature is dropping
%adiabatically as predicted by the analytic solution?

 %This is confirmed with numerical
%simulations of the plasma evolution \cite{ppr04PRA}.

%By combining fluorescence imaging and spectroscopy we have provided
%the first complete experimental characterization of the expansion
%dynamics of ultracold neutral plasmas. We have ...

We have demonstrated a plasma in which the expansion matches an
analytic solution to the Vlasov equations \cite{dse98} proposed as a
basic model for  the dynamics of quasineutral laser-produced
plasmas. To realize this situation experimentally, it is necessary
to create UNPs with a spherical Gaussian density distribution.
Relatively high initial electron temperature and low density are
also required to avoid collisional and radiative processes
\cite{kkb00,kon02,mck02,rha03} that would heat electrons or lead to
electron-ion equilibration. In future studies, deviations from the
model can be used to study these collisional processes
 when they become important.   The
 expansion dynamics shown here provide a general tool for developing a greater
understanding and intuition for the expansion of plasmas.

This work was supported by the National Science Foundation (Grant
PHY-0355069 and a grant for the Institute of Theoretical Atomic,
Molecular and Optical Physics (ITAMP) at Harvard University and
Smithsonian Astrophysical Observatory) and the David and Lucille
Packard Foundation.

%
%    \bibliographystyle{apsrev}
%    \bibliography{bibliography}

\end{document}